\documentclass[10pt, twocolumn]{article}
\pdfoutput=1
\usepackage{authblk}
\usepackage{amsmath}
\usepackage{graphicx}
\usepackage{float}
\usepackage{xcolor}
\usepackage{graphicx}
\usepackage{dcolumn}
\usepackage{bm}

\usepackage[T1]{fontenc}

\usepackage[left=1.50cm, right=1.50cm, top=1.50cm, bottom=2.50cm]{geometry}

\def \etal {{\it et al.}}
\title{Magnetic field effects on the proton EDM in a continuous all-electric storage ring}

\author[1]{Selcuk Hac{\i}{\" o}mero{\u g}lu}
\author[2]{Yuri F. Orlov}
\author[1,3]{Yannis K. Semertzidis}

\affil[1]{\small IBS, Center for Axion and Precision Physics, \textit{Daejeon, 34051, South Korea}}
\affil[2]{\small Cornell University, Department of Physics, \textit{Ithaca, New York 14853-2501, USA} }
\affil[3]{\small KAIST, Department of Physics, \textit{Daejeon, 34141, South Korea}}

\begin{document}

\onecolumn
\maketitle
\begin{abstract}
Electric dipole moment of the proton can be searched in an electric storage ring by measuring the spin precession rate of the proton beam on the vertical plane. In the ideal case, the spin precession comes from the coupling between the electric field and the electric dipole moment. In a realistic scenario, the magnetic field becomes a major systematic error source as it couples with the magnetic dipole moment in a similar way. The beam can see the magnetic field in various configurations which include direction, time dependence, etc. For instance, geometric phase effect is observed when the beam sees the field at different directions and phases periodically. We have simulated the effect of the magnetic field in the major independent scenarios and found consistent results with the analytical estimations regarding the static magnetic field cases. We have set a limit for the magnetic field in each scenario and proposed solutions to avoid systematic errors from magnetic fields.
\end{abstract}


\twocolumn

\section{Introduction}
Storage ring electric dipole moment (EDM) experiments \cite{ref:edm_in_sring, ref:edm_paper} aim to measure the vertical spin precession to see a coupling between the vertical spin component and the radial electric field, which should be proportional to the particle's EDM value. The proton EDM (pEDM) experiment is designed to be done with two simultaneously counter-rotating beams. The total EDM signal will be the sum of the two. 

According to the T-BMT equation \cite{ref:bmt}, the spin precession rate of the particle is
\begin{equation}
\small
\begin{split}
\vec \omega= \frac{e}{m}\vec s \times & \left[\frac{G \gamma+ 1}{\gamma} \vec B - \frac{G \gamma}{\gamma+1}\vec \beta (\vec \beta \cdot \vec B) \right. \\ 
&-\left(G+\frac{1}{\gamma+1} \right)\frac{\vec \beta \times \vec E}{c} \\
&+  \left.\frac{\eta}{2c}\left(\vec E -\frac{\gamma}{\gamma+1} \vec \beta (\vec \beta \cdot \vec E) + c \vec \beta \times \vec B \right) \right] ,
\end{split}
\label{eq:bmt}
\end{equation}
where $c$, $e$ and $m$ are the speed of light, the electric charge and the mass of the particle, $G=g/2-1$ is the magnetic anomaly ($\approx 1.8$ for proton),  $\vec \beta$ and $\gamma$ are the relativistic velocity and the Lorentz factor correspondingly, $\vec B$ and $\vec E$ are the magnetic and electric fields respectively. $\eta$ is the EDM coefficient.  In this work, it is taken as $\eta = 1.88 \times 10^{-15}$, corresponding to $d_p = 10 ^{-29}$ e$\cdot$cm.

As seen in the equation, the coupling between $\vec B$ and $G$ can contribute to the spin precession as a false EDM signal. We studied this effect separately for the ``static'' and ``alternating'' fields in the particle's rest frame. Each simulation was done with one proton particle in a continuous electric ring having a specific field index $m=0.2$, whose square root gives the vertical tune. The particle in each simulation has the so-called ``magic momentum'', which ideally freezes the average spin precession on the horizontal plane \cite{ref:frozen_spin_2, ref:frozen_spin_3}. For the protons, the magic momentum is 0.7 GeV/$c$.

\section{The EDM signal}
\begin{figure}[h]
	\includegraphics[width=\linewidth]{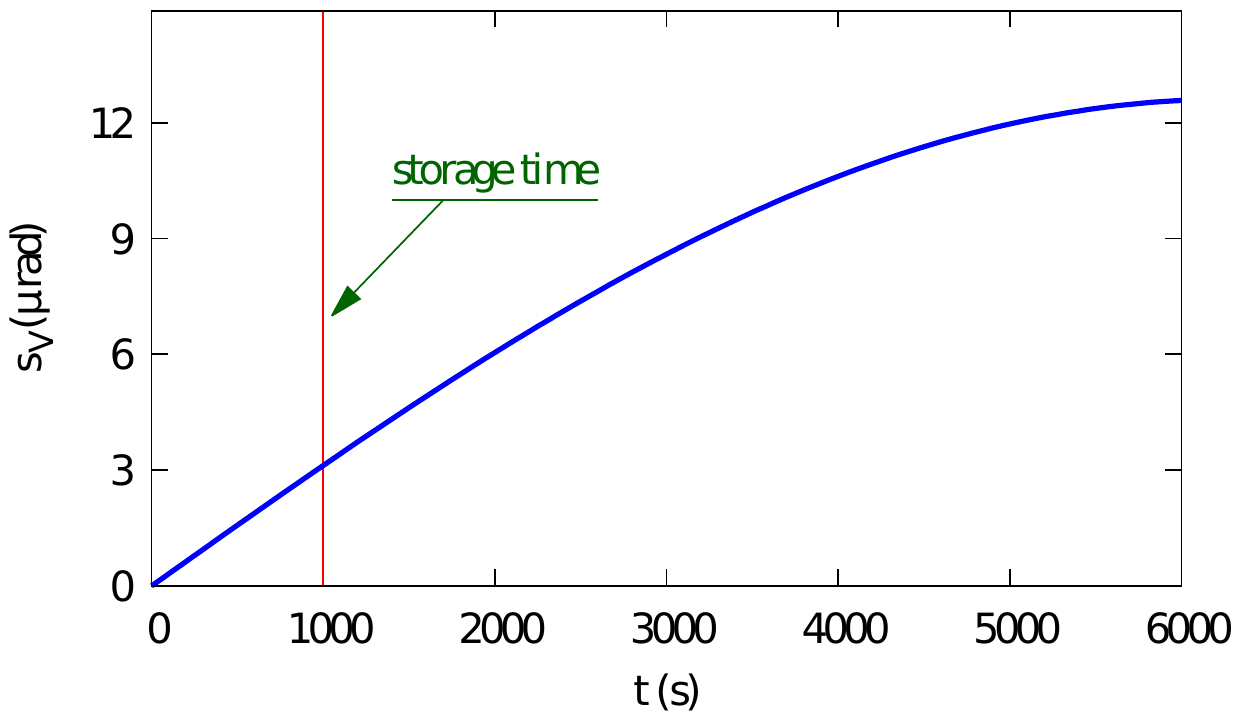}
	\caption{Vertical spin precession due to the radial electric field changes sinusoidally and gives about 3 nrad/s precession rate in the linear region. }
	\label{fig:sz_vs_t_E}
\end{figure}

In the absence of the magnetic and the longitudinal electric fields, the radial electric field couples with the EDM term to precess the spin on the vertical plane:
\begin{equation}
\omega_R = \frac{\eta e}{2 m c}  E_R s_L = \frac{\eta e}{2 m c} E_R \cos(\omega_a t),
\label{eq:dot_s_V_E}
\end{equation}
where $R$ and $L$ represent the radial and longitudinal directions respectively, $E_R$ is the radial electric field component, $t$ is time, and $\omega_a$ is the on-plane component of the spin precession rate. It can be kept small by means of a feedback RF cavity and sextupole fields \cite{ref:edm_in_sring, ref:edm_paper, ref:frozen_spin_2, ref:frozen_spin_3, ref:cosy_sct}. $s_L=\cos(\omega_a t)$ is a good approximation if the initial spin is totally longitudinal and the growth rate of the vertical spin component $s_V$ is negligible. Then, $s_V$ becomes
\begin{equation}
s_V =  \frac{\eta e}{2 m c} E_R  \int_{0}^{t}\cos(\omega_a t) dt = \frac{\eta e}{2 m c \omega_a} E_R \sin(\omega_a t).
\label{eq:s_V_E}
\end{equation}
Figure \ref{fig:sz_vs_t_E} shows the Equation \ref{eq:s_V_E} for $\omega_a = 0.25$ cycles/s and $E_R=10$ MV/m. For a small $\omega_a t$, the vertical spin component should be
\begin{equation}
s_V=\frac{\eta e}{2 m c} E_R t \approx 3 \times 10^{-9}t.
\end{equation}

\section{Effect of the static magnetic fields}
Disregarding the EDM term, Equation \ref{eq:bmt} around the radial axis becomes
\begin{equation}
\begin{split}
\omega_R=\frac{e}{m} &\Bigg[\left( G+\frac{1}{\gamma} - \frac{G\gamma \beta^2}{\gamma+1} \right) B_L s_R  + \\ & \Bigg( \left(G+\frac{1}{\gamma} \right)B_R - \left(G+\frac{1}{\gamma+1}\right)\frac{\beta_L E_V}{c} \Bigg)s_L \Bigg].
\end{split}
\label{eq:bmt_ver_spin_prec}
\end{equation}
The radial spin component ($s_R$) couples with a longitudinal magnetic field ($B_L$). On the other hand, the longitudinal spin component ($s_L$) couples with the radial magnetic field ($B_R$) and the vertical electric field ($E_V$), which averages to zero if $B_R=0$ (because of the electrical focusing). However, nonzero $B_R$ introduces a vertical offset and the quadrupole component of the electric field applies vertical force on the beam to balance it.

As seen, the beam dynamics imposes a unique effect from each directional magnetic field. Therefore, they will be shown separately.

\subsection{Radial magnetic field}

If the magnetic field is purely radial, Equation \ref{eq:bmt_ver_spin_prec} simplifies to 
\begin{equation}
\omega_R=\frac{e s_L}{m} \left[\left(G+\frac{1}{\gamma} \right)B_R - \left(G+\frac{1}{\gamma+1}\right)\frac{\beta_L E_V}{c} \right].
\label{eq:bmt_rad}
\end{equation}
$B_R$ also moves the beam vertically by
\begin{equation}
y=-\frac{\beta c R_0 B_R}{E_R Q_y^2},
\label{eq:y_due_to_Br}
\end{equation}
where $R_0$ is the radius of the ring and $E_R$ is the radial electric field on the particle. On the other hand, the vertical electric field due to the weak electric focusing is
\begin{equation}
E_V\approx -E_R \frac{n y}{R_0},
\label{eq:E_V_basic}
\end{equation}
with $n=m+1$. Combining Equations \ref{eq:y_due_to_Br} and \ref{eq:E_V_basic}, one obtains
\begin{equation}
E_V \approx \frac{E_R Q_y^2}{R_0}\frac{\beta c R_0 B_R}{E_R Q_y^2}=\beta c B_R.
\end{equation}
Then, using the identity $\beta^2=1-1/\gamma^2$ and the definition of magic momentum 
\begin{equation}
	G=1/(\gamma^2-1),
	\label{eq:magic_mom}
\end{equation}
the parenthesis in Equation \ref{eq:bmt_rad} simplifies to
\begin{equation}
\begin{split}
	\omega_R &= \frac{e s_L}{m}\left( G + \frac{1}{\gamma} - G \beta^2 - \frac{\beta^2}{\gamma+1}  \right)B_R \\ &= \frac{e s_L}{m}G B_R .
\end{split}
	\label{eq:wr_from_Br}
\end{equation}

\begin{figure}
	\centering
	\includegraphics[width= \linewidth]{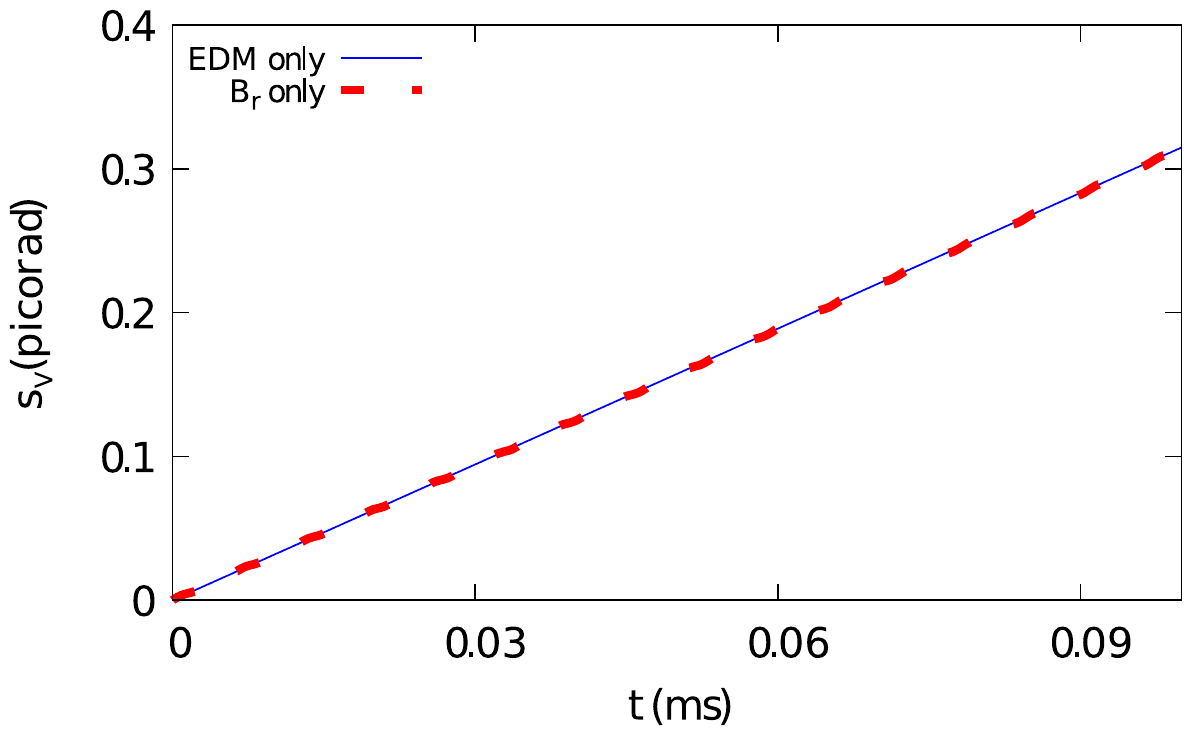}
	\caption{Simulations with $E_R=10$ MV/m coupling with the EDM and $B_R=16.7$ aT give the same spin precession as estimated analytically. The wiggles in the $B_R$ case originate from betatron oscillations. $\eta=1.88\times 10^{-15}$ in the simulations.}
	\label{fig:sy_br_and_er}
\end{figure}

That means, the radial magnetic field coupling with $G$ mimics the radial electric field coupling with $\eta /2 c$ (See Equation \ref{eq:bmt}). Comparing these two terms one finds that $B_R=16.7$ aT and $E_R=10$ MV/m lead to the same vertical spin precession rate for $\eta=1.88 \times 10^{-15}$ as shown in Figure \ref{fig:sy_br_and_er}.

Alternatively, the same result can be obtained following the Lorentz equation:
\begin{equation}
	\vec F_y = e(\vec E_V + \vec \beta \times \vec B_R)=0 \quad \longrightarrow \quad \vec E_V=-\vec \beta \times \vec B_R.
\end{equation}
In the particle's rest frame, the magnetic field becomes
\begin{equation}
	\vec B'=\gamma (\vec B_R-\vec \beta\times \vec E_V)= \gamma \left(\vec B_R + \vec \beta \times (\vec \beta \times \vec B_R) \right).
\end{equation}
Eventually for the radial field one gets
\begin{equation}
	B' = \gamma (1-\beta^2)B_R = \frac{B_R}{\gamma}.
\end{equation}
In the rest frame of the particle, time slows down by a factor of $\gamma$. Then, for a longitudinally polarized beam
\begin{equation}
	\omega_R = \frac{eg}{2m}<B'> = \frac{eg}{2m\gamma^2}B_R.
	\label{eq:wr_from_Br_2}
\end{equation}
For the magic momentum, the comparison of Equations \ref{eq:wr_from_Br} and \ref{eq:wr_from_Br_2} yields 
\begin{equation}
	G=\frac{g}{2 \gamma^2}.
\end{equation}
This can also be shown with Equation \ref{eq:magic_mom} using the identity $G=g/2-1$:
\begin{equation}
	\frac{g}{2\gamma^2}=\frac{g/2}{1+\frac{1}{G}}=\frac{G+1}{\frac{G+1}{G}}=G.
\end{equation}

\begin{figure}
	\centering
	\includegraphics[width=0.5\linewidth]{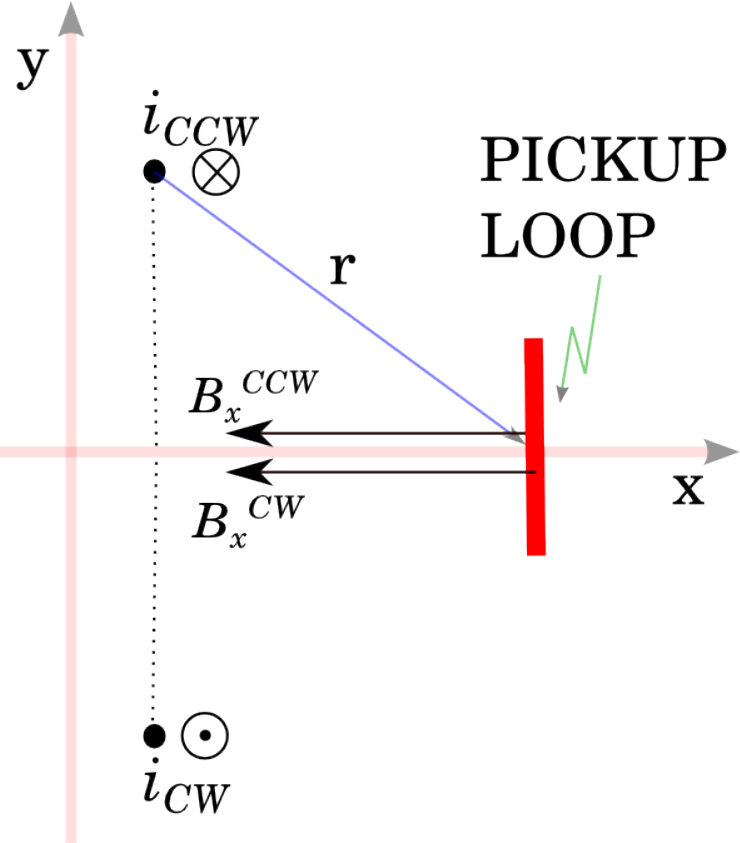}
	\caption{The split counter-rotating beams induce a net magnetic field on the horizontal plane. CW: Clockwise, CCW: Counter-clockwise}
	\label{fig:cw_ccw_b_field}
\end{figure}

$B_R$ splits the counter-rotating beams vertically according to Equation \ref{eq:y_due_to_Br}. Then, the  split beams induce a net magnetic field on the horizontal plane (Figure \ref{fig:cw_ccw_b_field}). The induced magnetic field is of the order of attoTelsa for the above-mentioned conditions. The pEDM collaboration proposes a long-term measurement with SQUID-based BPMs and cancellation with Helmholtz coils \cite{ref:edm_paper}.

\subsection{Longitudinal and vertical magnetic fields}
\begin{figure}[h]
	\centering
	\includegraphics[width=0.6\linewidth]{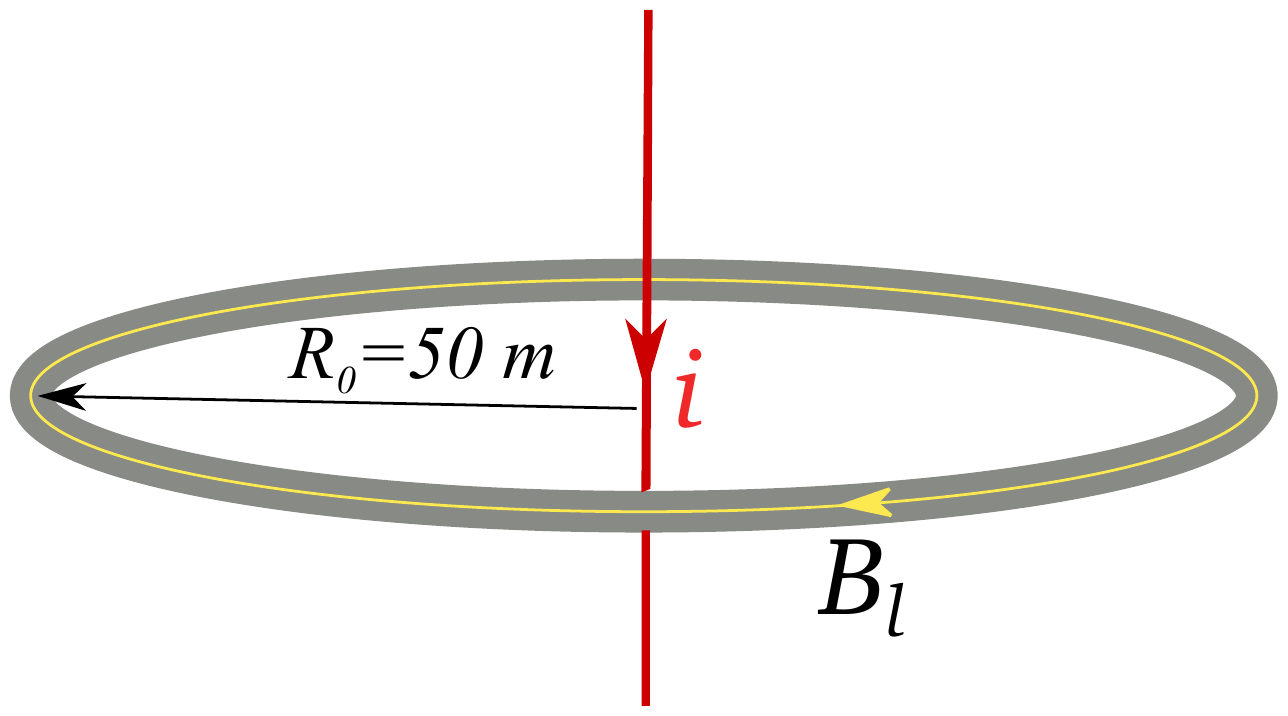}
	\caption{ $i=25$ mA DC current at the center of the ring induces approximately $B_l=1$ nT static longitudinal magnetic field along the beam path.}
	\label{fig:static_long_b_field}
\end{figure}

A 1nT static longitudinal magnetic field can be generated by a 25 mA current passing through the center of a 300m long ring as shown in Figure \ref{fig:static_long_b_field}.

In the absence of the EDM and $B_R$ terms, $\omega_R$ of Equation \ref{eq:bmt_ver_spin_prec} becomes
\begin{equation}
\begin{split}
\omega_R&=\frac{e}{m}\left[ G+\frac{1}{\gamma} - \left(\frac{G\gamma}{\gamma+1}\right)\beta^2 \right]B_L s_R \\
& = \frac{e}{m}\frac{g}{2\gamma}B_Ls_R.
\end{split}
\label{eq:y_due_to_Bl}
\end{equation}
Due to the context of the study, again we assume a small vertical spin component. In that case, $s_L$ and $s_R$ change sinusoidally. Assuming a longitudinal spin at the beginning of the storage, the longitudinal and the radial spin components are given as
\begin{equation}
\begin{split}
& s_L (t) = \cos(\omega_a t) ,\\
& s_R (t) =\sin(\omega_a t).
\label{eq:sl_sr}
\end{split}
\end{equation}
Even at the magic momentum, the horizontal spin component of the particle grows by $\omega_a$, if there is a vertical magnetic field. The longitudinal magnetic field couples with the vertical spin component and leads to some nonzero $\omega_a$ too, but this is about 6 orders of magnitude smaller than a comparable vertical magnetic field would cause. Beside these, the momentum spread contributes to $\omega_a$ as well, by an amount of <1 mrad/s as a requirement in the pEDM experiment \cite{ref:edm_paper}.

\begin{figure}
	\centering
	\includegraphics[width=\linewidth]{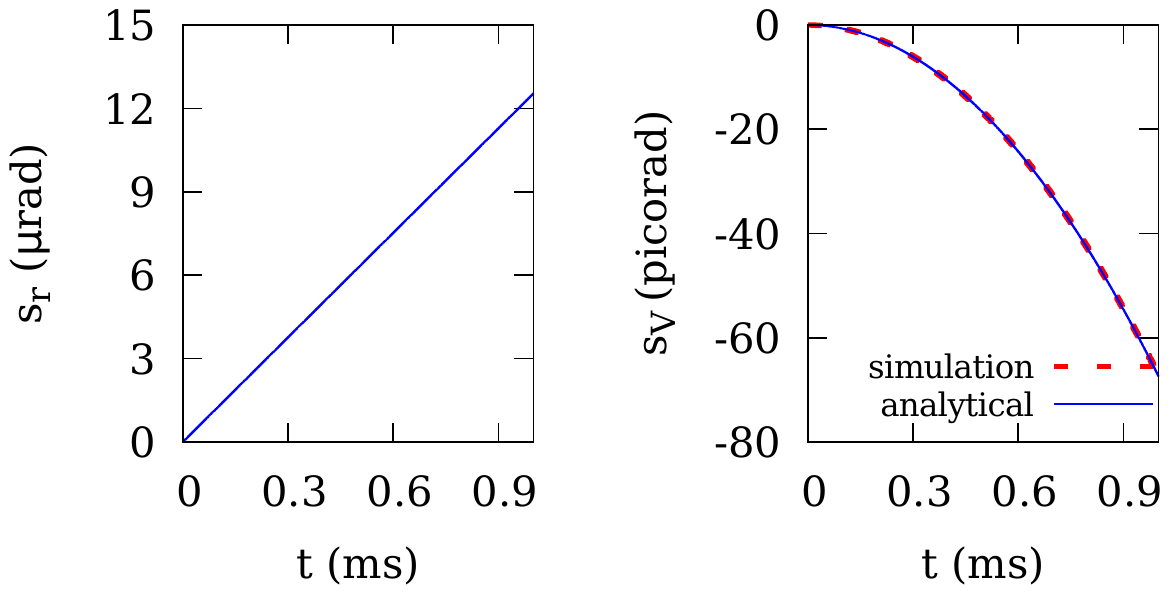}
	\caption{The spin components as simulated for 1 ms storage time with a magic particle in an electric ring. Because of the short storage time compared to one cycle of $\omega_a$, $s_R$ changes linearly, and $s_V$ approximates to a quadratic function (See Equations \ref{eq:sl_sr} and \ref{eq:s_V}). \textit{Left}: 50pT vertical magnetic field causes $\omega_a \approx 12.5$ mrad/s on the horizontal plane. \textit{Right}: Having linear dependence on $\omega_a$, $s_V$ has quadratic dependence on time.  Combination of 50pT longitudinal and vertical static magnetic fields grows the vertical spin component up to 67 prad , matching well with the analytical estimation.}
	\label{fig:sr_and_sy}
\end{figure}

Defining $k \equiv eg/2m\gamma=2.2 \times 10^8$ C/kg for the proton at the magic momentum, the integral of Equation \ref{eq:y_due_to_Bl} becomes
\begin{equation}
\begin{split}
s_V(t) &= -k B_L \int_{0}^{t} \sin(\omega_a t) dt = \frac{k B_L}{\omega_a}\cos(\omega_a t) \Big \vert_{0}^{t} \\ 
&= \frac{k B_L}{\omega_a}  \Big[\cos(\omega_a t) -1\Big].
\label{eq:s_V}
\end{split}
\end{equation}
For small $\omega_at$ values, $s_R=\sin(\omega_a t) \approx \omega_a t$. Then, the slope of $s_R$ vs. $t$ plot is equal to $\omega_a$, which is determined by the ring design, particle momentum and the vertical magnetic field. The simulation with a 50 pT vertical magnetic field yields $\omega_a \approx 12.5$ mrad/s (left plot of Figure \ref{fig:sr_and_sy}). This is quite a fast on-plane precession, but the conclusions from this 1 ms simulation hold for a smaller $\omega_a$ as well. Substituting the values in Equation \ref{eq:s_V} gives
\begin{equation}
\begin{split}
s_V&\approx \frac{2.2 \times 10^{8} \times 50 \times 10^{-12}}{\omega_a} \times \Big[-\frac{(\omega_a t)^2}{2}\Big] \\
&= -5.5 \times 10^{-3} \omega_a t^2.
\label{eq:sy_due_to_Bl_final}
\end{split}
\end{equation}
 Note that the vertical magnetic field does not affect $s_V$ directly, but it has an indirect effect via $\omega_a$.
Figure \ref{fig:sr_and_sy} shows a good agreement between the estimation  at Equation \ref{eq:sy_due_to_Bl_final} and the Runge-Kutta simulation. The details of the simulation tool can be found at \cite{ref:selcuk_rk4}.

A comparison with Figure \ref{fig:sz_vs_t_E} shows that the pT level of static longitudinal and vertical magnetic fields lead to a nine orders of magnitude larger $s_V$ than the EDM signal at the end of the storage. 

\subsubsection{Eliminating the effect of the longitudinal magnetic field}
One needs to have a 1 fT level average magnetic field in vertical and longitudinal directions to reduce the effect to the level of the EDM signal (nrad/s).

As Equation \ref{eq:y_due_to_Bl} shows, the effect of the $B_L$ amplifies proportionally with $s_R$. This effect can be exploited by using a radially polarized test bunch. According to Equation \ref{eq:y_due_to_Bl}, the spin precession rate from $B_L=1$ fT is $\omega_R=2.2\times 10^8 \times 10^{-15}=220$ nrad/s without any contribution from the EDM as $\vec s \times \vec E=0$ Monitoring that bunch with the polarimeter \cite{ref:polarimeter}, its $\omega_R$ can be frozen by applying an inverse longitudinal magnetic field with 1 fT resolution.

Moreover, the $90^\circ$ phase difference between the EDM signal and this false one (compare the Equations \ref{eq:s_V_E} and \ref{eq:s_V}  or Figures \ref{fig:sz_vs_t_E} and \ref{fig:sr_and_sy}) can be exploited to further filter out the effect of $B_L$ offline.

\section{Effect of alternating magnetic field and the geometric phase effect}
\begin{figure}[h!]
	\includegraphics[width=\linewidth]{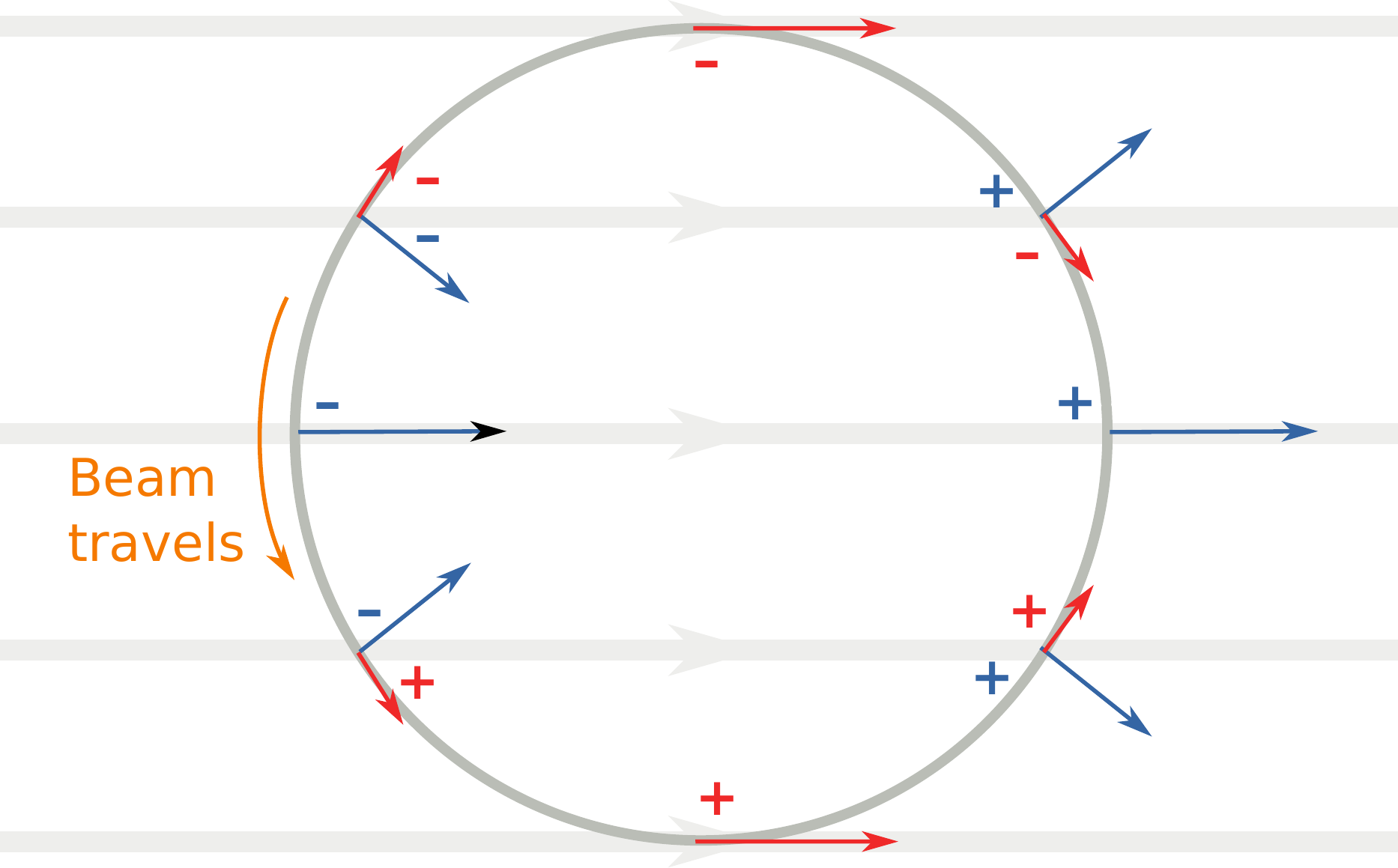}
	\caption{While being static in the lab frame, earth's field is seen as an alternating field in the particle's rest frame.  This kind of a configuration may lead to nonzero $s_V$ even though the average field along the ring is exactly zero. "$+$" and "$-$" signs show the direction of the field as the particle sees it.}
	\label{fig:geometric_phase_demo}
\end{figure}

In this section, alternating field refers to the particle's rest frame and no time dependence in the lab frame is considered. Alternating magnetic field can show up in a number of ways. Figure \ref{fig:geometric_phase_demo} shows one case originating from the earth's DC field. 

\begin{figure}[h!]
	\includegraphics[width=\linewidth]{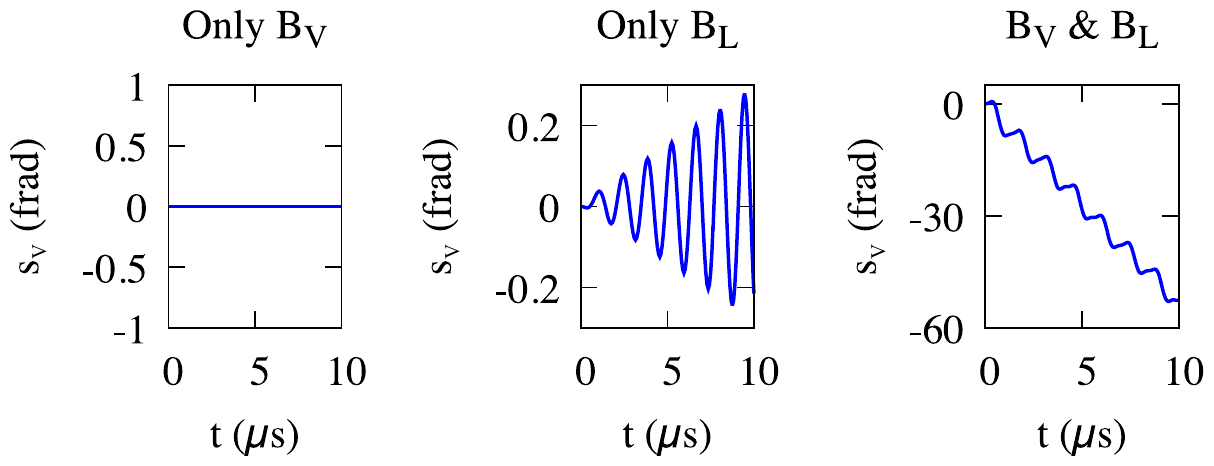}
	\caption{$s_V$ for various cases of alternating vertical ($B_V$) and longitudinal ($B_L$) sinusoidal magnetic fields. In each case, the field makes one oscillation around the ring. The non-magic particle had a nonzero horizontal spin precession rate ($\omega_a$) in the simulations. \textit{Left}: As expected, the vertical magnetic field alone does not lead to any vertical spin precession. \textit{Middle}: Longitudinal magnetic field couples with the increasing radial spin component to increase the oscillation amplitude of $s_V$. Note that $<s_V>=0$ since $<B_L>=0$. \textit{Right}: Combination of the vertical and the longitudinal magnetic fields enhances the vertical spin precession, which in some cases does not average to zero. }
	\label{fig:sz_for_N_1_demo}
\end{figure}

At first glance one may think that sinusoidal magnetic field along the lattice averages to zero with no spin growth effect. However some magnetic field configurations can make $s_V$ grow with time. This is closely related to the geometric phase effect \cite{ref:berry_geom_phase, ref:deuteron_ags}. It was previously pointed out as a systematic error for storage ring EDM experiments \cite{ref:y_orlov_geom_phase} as well as for neutron EDM experiments \cite{ref:geom_phase_nEDM}. Figure \ref{fig:sz_for_N_1_demo} shows simulation results with vertical and longitudinal magnetic field combinations as a demonstration.

This section aims to show the effect of the alternating magnetic field configurations. All major independent configurations of magnetic field were studied in this section, including combinations of two perpendicular  directions ($B_L\&B_V$, $B_L\&B_R$ and $B_R\&B_V$) with 0 and 90 degree phase differences. The magnetic field in a particular direction changes sinusoidally to make one cycle along the ring ($N=1$).

\begin{figure}[h!]
	\centering
	\includegraphics[width=\linewidth]{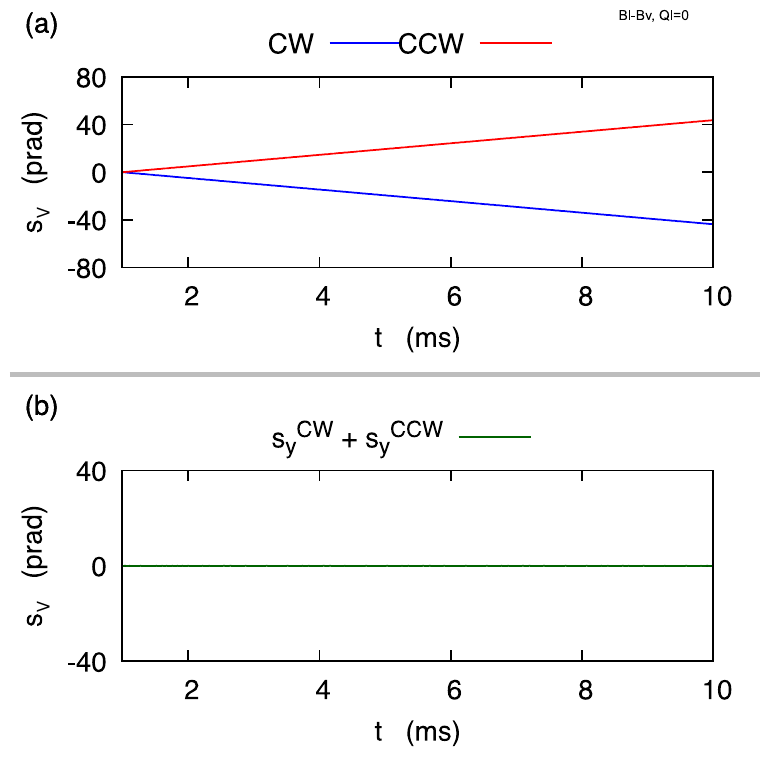}
	\caption{Running average of the vertical spin precession due to the combination of 1 nT amplitude vertical and longitudinal magnetic fields with $90^\circ$ phase difference. The plots include vertical offsets for legibility. (a) shows the running average of the counter-rotating particles individually. Again, the fields oscillate by one turn per cycle ($N=1$). The spin precession rate of an individual particle is much faster than the EDM signal.  However, the spin direction for the counter-rotating particles is opposite to the EDM, leading to a cancellation of $s_V$ in total (b).}
	\label{fig:sy_V90_L0}
\end{figure}

The  notion of counter-rotating beams introduces an additional symmetry and is useful for eliminating many systematic errors including the geometric phase effect. For instance combination of alternating 1nT longitudinal and vertical magnetic fields (of $N=1$) with $90^\circ$ phase difference leads to $s_V\approx 40$ nrad/s, much larger than the EDM effect  (See Figure \ref{fig:sy_V90_L0}). But unlike the EDM effect, this $s_V$ has an opposite sign for the clockwise (CW) and the counter-clockwise (CCW) beams with the sum cancelling out. 

The simulation results show several qualitatively different effects including the one shown in Figure \ref{fig:sy_V90_L0}. The magnetic field directions and the phase difference ($\phi$) between them determines which class the effect falls into. It is also seen that the effect in each case either cancels or averages out to <10 prad/s after 10 ms for 1 nT field values. 

\begin{figure}[h!]
\centering
\includegraphics[width=\linewidth]{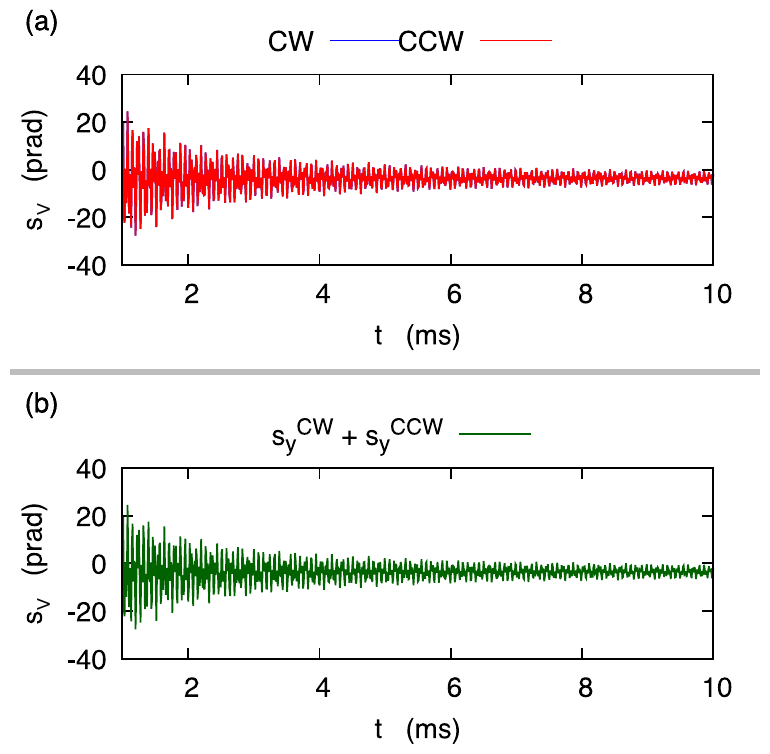}
\caption{Running average of $s_V$ of CW and CCW particles (\textit{a}) and their sum (\textit{b}). Amplitudes of $B_R$ and $B_L$ are 1nT with $N=1$. The angle between $B_R$ and $B_L$ does not make a difference. Combination of $B_R$ and $B_V$ with $90^\circ$ phase difference also gives a similar $s_V$. In this class, $s_V$ of CW and CCW particles have the same value. Therefore they don't cancel each other. But their running averages individually go below the EDM limit. For instance in this simulation the total $s_V$ has a slope less than 10 prad/s at the last 5 ms of the tracking.}
\label{fig:sy_L0_R0}
\end{figure}

\begin{figure}[h!]
\centering
\includegraphics[width=\linewidth]{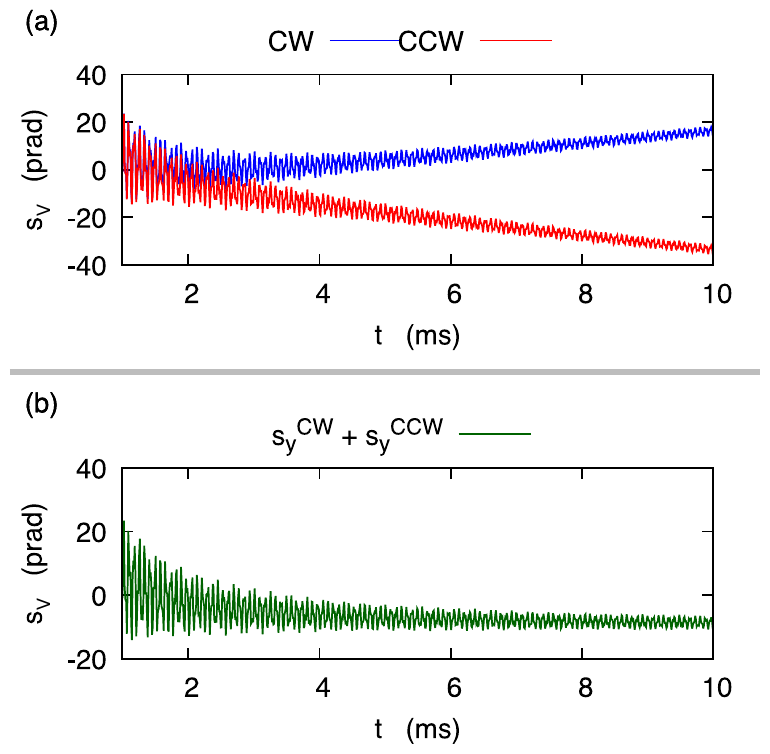}
\caption{The running average of $s_V$ as obtained in the presence of radial and vertical magnetic fields of the same phase. The $s_V$ of CW and CCW are in the opposite direction (\textit{a}) but they don't cancel exactly (\textit{b}). The total effect averages out similarly as in Figure \ref{fig:sy_L0_R0}.}
\label{fig:sy_V0_R0}
\end{figure}

Figures \ref{fig:sy_L0_R0} and \ref{fig:sy_V0_R0} show two classes obtained with radial magnetic field. The alternating radial magnetic field can make $s_V$ alternate at each cycle. The effect averages out in Figure \ref{fig:sy_L0_R0}, but not in Figure \ref{fig:sy_V0_R0}. Coupling with a vertical magnetic field of the same phase makes $s_V$ grow with time. This originates from the fact that the radial spin component is not the same when the radial  magnetic field increases and decreases. Still, the effect cancels out to first order when the signal from CW and CCW are added up. In each case, the average $s_V$ decreases to <10 prad/s after 10 ms.

\begin{figure}[H]
	\centering
	\includegraphics[width=0.5\linewidth]{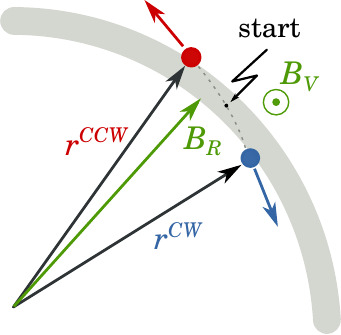}
	\caption{Vertical magnetic field makes the CW and CCW particles shift to slightly different radial positions. This causes a change in the momentum, hence the radial spin component.}
	\label{fig:b_ver_split}
\end{figure}

It is worth mentioning that in Figure \ref{fig:sy_V0_R0}, $s_V$ of CW and CCW particles don't cancel out exactly despite their symmetric appearance. The vertical magnetic field makes the CW and CCW particles move on slightly different radial positions in the ring and have different radial spin components $s_R$ (Figure \ref{fig:b_ver_split}). This eventually causes a phase difference between $s_V$ of CW and CCW particles, even though the total spin precession rate averages out to negligible levels.

\begin{figure}
	\centering
	\includegraphics[width=\linewidth]{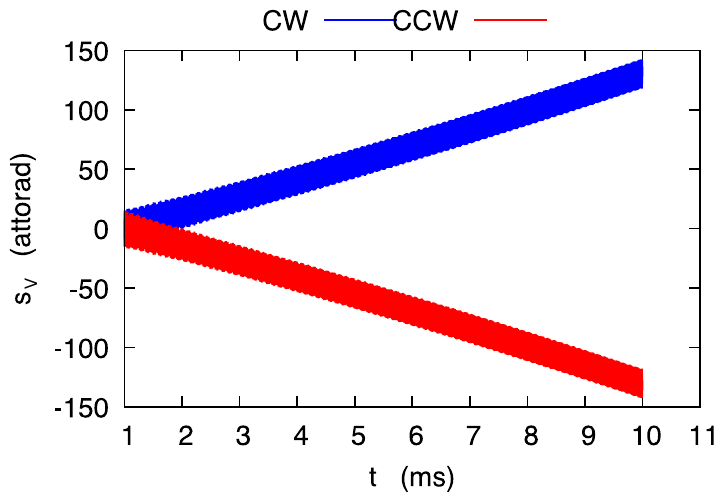}
	\caption{The running average of $s_V$ as obtained in the presence of longitudinal and vertical magnetic fields of the same phase. The offset is shifted in the plots for legibility. The total field is cancelled to atto-radian level. Yet this effect is more likely a numerical error rather than being physical.}
	\label{fig:sy_V0_L0}
\end{figure}

Finally, the simulations with vertical and longitudinal magnetic fields of the same phase yielded $\omega_r\approx 15$ frad/s as seen in Figure \ref{fig:sy_V0_L0}. While one can argue that the effect is too small and the CW/CCW cancel each other, we observed that unlike the other cases, $\omega_r$ depends strongly on the time steps of the simulations. Therefore it seems more like a numerical error.

\section{Summary of the magnetic field scenarios}
The static and alternating magnetic fields (in the particle's rest frame) require different treatment. The static magnetic fields can be actively cancelled by continuous measurement and feedback. The most sensitive measurement should be made on the radial direction. The vertical magnetic field does not affect the vertical spin component directly, but it enhances the effect of the longitudinal field component, which can be avoided by the help of a test bunch of $90^\circ$ polarization at every injection. Also, the vertical magnetic field can be eliminated by using a BPM similar to the case of radial magnetic field.

According to the presented simulations, a 1nT amplitude of field grows the vertical spin component by less than 10 prad/s rate, two orders of magnitude less than the EDM signal. The magnetic field can be shielded to this level within the present technology \cite{ref:fierlinger}. These results include the geometric phase effects as well. 

Table \ref{tbl:summary} summarizes the scenarios causing the spin precession rate on the vertical plane and the proposed solutions. The numbers are estimated for an all-electric ring of 50 m radius.

\begin{table}
	\caption{Summary of the major independent  magnetic field configurations.  $\langle \omega_r \rangle$ is the average spin precession rate. Each simulation was done with 1 nT magnetic field strength.}
\begin{tabular}{p{1.3cm}p{0.8cm}p{2cm} p{3.4cm}}
	\hline
	\label{tbl:summary}
	Field & AC Phase& $\langle \omega_r \rangle$ [rad/s] & Solution \\
	\hline
	DC $B_R$ &n/a& 0.18 & Measurement and active cancellation with BPMs \\
	DC $B_L$ &n/a&$<5.5\times 10^{-6}$, proportional to $\omega_a$& Current to be limited to $<1$mA and DC $B_V$  to be avoided \\ 
	DC $B_V$&n/a& 0 & Can be avoided with BPM similar to $B_R$ case \\
	$B_V\&B_L$ &$90^\circ$&$9 \times 10^{-9}$ & CW/CCW cancel\\
	$B_R\&B_V$ &$0^\circ$ & $3.5 \times 10^{-9}$ & CW/CCW average out \\
	$B_R\&B_L$ &$0, 90^\circ$ & $<10^{-10}$ & CW/CCW average out \\
	$B_R\&B_V$ &$90^\circ$ & $<10^{-10}$ & CW/CCW average out \\
	$B_V\&B_L$ &$0^\circ$&Negligible & \\
	\hline
\end{tabular}
\end{table}

\section{Conclusions}

The storage ring proton EDM experiment has several features to control the spin precession. These features either cancel the false EDM signal directly, or help identifying the direction of the magnetic field for active cancellation. 

An important instrument in the experiment is the simulatenous storage of the counter-rotating beams. The symmetry of their motion either makes their spins precess in the cancelling directions or helps identifying the field.

The control over the beam polarization is another strong feature, because the spin of the beam is insensitive to the magnetic field in that direction. This helps identifying the magnetic field of a particular direction.

Studying the major independent magnetic field scenarios, we have seen that the related systematic errors can be kept under control with the tools using the present technology.

\section{Acknowledgement}
IBS-Korea (project system code: IBS-R017-D1-2018-a00) supported this project.

\onecolumn

\end{document}